\documentclass[final]{svjour3}

\usepackage{graphicx}
\usepackage{rotating}
\usepackage{amssymb}
\usepackage{mathptmx}
\makeatletter
\journalname{Journal of Low Temperature Physics}

\newcommand{\be}{\begin{equation}}
\newcommand{\ee}{\end{equation}}

\newcommand{\dlt}{\delta}
\newcommand{\prt}{\partial}
\newcommand{\br}{{\bf r}}

\newcommand{\al}{\alpha}

\newcommand{\om}{\omega}

\newcommand{\rgl}{\rangle}\newcommand{\lgl}{\langle}

\begin{document}

\newcommand{\hdblarrow}{H\makebox[0.9ex][l]{$\downdownarrows$}-}

\title{Strongly nonequilibrium Bose-condensed atomic systems}

\author{V.I. Yukalov$^{1,2}$ \and A.N. Novikov$^{1,2}$ \and \\ V.S. Bagnato$^1$}

\institute{1:Instituto de Fisica de S\~{a}o Calros, Universidade de S\~{a}o Paulo,\\
CP 369, 13560-970 S\~{a}o Carlos, S\~{a}o Paulo, Brazil\\
Tel.: +7 (496) 21 63947 \\ Fax: +7 (496) 21 65084\\
\email{yukalov@theor.jinr.ru}
\\2: Bogolubov Laboratory of Theoretical Physics,\\
Joint Institute for Nuclear Research, Dubna 141980, Russia}

\date{XX.XX.2014}

\maketitle

\keywords{trapped atoms, Bose-Einstein condensate, nonequilibrium states,
vortex turbulence, grain turbulence, wave turbulence}

\begin{abstract}

A trapped Bose-Einstein condensate, being strongly perturbed, exhibits
several spatial structures. First, there appear quantum vortices.
Increasing the amount of the injected energy leads to the formation of
vortex tangles representing quantum vortex turbulence. Continuing energy
injection makes the system so strongly perturbed that vortices become 
destroyed and there develops another kind of spatial structures with 
essentially heterogeneous spatial density. These structures consist of 
high-density droplets, or grains, surrounded by the regions of low density. 
The droplets are randomly distributed in space, where they can move; 
however they live sufficiently long time to be treated as a type of 
metastable creatures. Such structures have been observed in nonequilibrium 
trapped Bose gases of $^{87}$Rb subject to the action of an oscillatory 
perturbation modulating the trapping potential. Perturbing the system even 
stronger transforms the droplet structure into wave turbulence, where Bose 
condensate is destroyed. Numerical simulations are in good agreement with 
experimental observations.

PACS numbers: 03.75.-b, 03.75.Hh, 05.30.Jp, 47.27.E-, 47.27.Gs, 67.40.Vs
\end{abstract}

\section{Introduction}

Bose atoms in a trap can be cooled down to such low temperatures when almost
all of them pile down to the ground state corresponding to Bose-Einstein
condensate. If the trapped Bose-condensed cloud is slightly perturbed, it
exhibits elementary collective excitations, similarly to other finite systems,
such as quantum dots, atomic nuclei, or clusters
\cite{Reimann_1,Yannouleas_2,Cejnar_3,Saarikoski_4,Birman_5}. But what
happens when the Bose-condensed cloud is strongly perturbed?

We know that, when the external perturbation imposes a rotation moment, there
can arise quantum vortices that may form a vortex lattice
\cite{Cooper_6,Fetter_7}. But if the external perturbation imposes no rotation
moment, but just strongly shakes the condensate, as has been suggested in
Refs. \cite{Yukalov_8,Yukalov_9}, then no vortex lattices arise. Instead, the
increasing number of vortices forms a random tangle corresponding to quantum
turbulence \cite{Feynman_10,Tough_11,Tsubota_12,Tsubota_13}. This vortex
turbulence of trapped atoms has been observed in three-dimensional traps
\cite{Henn_14,Shiozaki_15,Seman_16} as well as in quasi-two-dimensional traps
\cite{Wilson_17}, and is summarized in the review articles
\cite{Tsubota_18,Nemirovskii_19,Bagnato_20}.

In the present paper, we study what could be other possible nonequilibrium
states in a trapped strongly perturbed Bose-Einstein condensate. We investigate
this problem both in experiments with trapped atoms of $^{87}$Rb and
accomplishing computer simulations for the setup exactly corresponding to these
experiments. Observed experimental data are in good agreement with numerical
simulations.

\section{Methods of strong perturbation}

Generally, a Bose-Einstein condensate is characterized by an equation for
the condensate wave function $\eta = \eta({\bf r},t)$. For a system of weakly
interacting atoms at temperatures close to zero, the condensate-function
equation takes the form of the nonlinear Schr\"{o}dinger equation (NLS)
\be
\label{1}
 i\; \frac{\prt}{\prt t} \; \eta = \left ( - \; \frac{\nabla^2}{2m} + U +
\Phi | \; \eta \; |^2 \right ) \eta \;  .
\ee
Here and in what follows, the Planck constant is set to one, $U = U({\bf r},t)$
is the total external potential, and the atomic interaction strength is
\be
\label{2}
 \Phi = 4\pi \; \frac{a_s}{m} \;  ,
\ee
with $a_s$ being scattering length and $m$, atomic mass.

When the system is in equilibrium, the condensate is in its ground state.
To perturb the condensate, it is necessary to inject energy into the system.
The amount of injected energy, per atom, during the excitation time $t$, 
can be represented as
\be
\label{3}
 E_{inj} = \frac{1}{N} \; \int_0^t \left | \left \lgl  \frac{\prt \hat H}{\prt t}
\right \rgl \right |
\; dt \; ,
\ee
where $N$ is the total number of atoms, and the energy Hamiltonian is
\be
\label{4}
  \hat H = \int \eta(\br, t) \left ( - \; \frac{\nabla^2}{2m} + U \right )
\eta(\br,t)\; d\br + \frac{1}{2} \; \Phi \int |\; \eta(\br,t) \; |^4 \; d\br \; .
\ee

This tells us that energy can be injected through one of two ways, either
adding to the trapping potential $U(\bf r)$ a time-dependent perturbative term,
so that the total external potential be the sum
\be
\label{5}
 U(\br,t) = U(\br) + V(\br,t) \;  ,
\ee
or by varying the scattering length, so that the interaction strength be
time-dependent,
\be
\label{6}
 \Phi(t) = 4\pi \; \frac{a_s(t)}{m} \;  .
\ee
The perturbations are assumed to be organized in such a way that no rotation
moment is imposed onto the system.

If the trap modulation is done by means of an alternating potential
$$
 V(\br,t) \sim A \cos(\om t) \qquad ( A \ge 0 ) \;  ,
$$
as is advanced in Refs. \cite{Yukalov_8,Yukalov_9}, then the injected energy, 
for sufficiently long time $t \gg 2\pi/\omega$, takes the form 
\be
\label{7}
 E_{inj} \sim \frac{2}{\pi} \; A \om t \;  ,
\ee
which makes it possible to study the amplitude-time phase diagram through 
the relation
\be
\label{8}
 A \sim \frac{\pi E_{inj}}{2 \om t} \;  .
\ee

The other possibility of injecting energy is by modulating the scattering
length, making it time-dependent with the help of Feshbach resonance
techniques, as has been suggested in Refs. \cite{Ramos_21,Yukalov_22,Yukalov_23}.
The time dependence of the scattering length is realized through the oscillating
magnetic field, when
\be
\label{9}
 a_s(t) = \overline a_s \left [ 1 - \; \frac{\Delta B}{B(t)-B_\infty}
\right ] \;   ,
\ee
where $\bar{a}_s$ is a background scattering length, $\Delta B$, resonance
width, and $B_\infty$ is a resonance field. The modulating magnetic field
oscillates by the law
\be
\label{10}
 B(t) = B_0 + B_1 \cos(\om t) \;   .
\ee
For a small oscillation amplitude, such that $|B_1/B_0| \ll 1$, the effective
scattering length is
\be
\label{11}
 a_s(t) \simeq a_0 + a_1 \cos(\om t) \;  ,
\ee
with the notation
$$
 a_0 \equiv \overline a_s \left (  1 - \; \frac{\Delta B}{B_0-B_\infty}
\right ) \; , \qquad
 a_1 \equiv \overline a_s \; \frac{B_1\Delta B}{(B_0-B_\infty)^2} \; .
$$
Then the interaction strength takes the form
\be
\label{12}
 \Phi(t)= \Phi_0 + \Phi_1 \cos(\om t) \;  ,
\ee
in which
$$
\Phi_0 \equiv 4\pi \; \frac{a_0}{m} \; , \qquad
\Phi_1 \equiv 4\pi \; \frac{a_1}{m} \;   .
$$
The injected energy reads as
\be
\label{13}
 E_{inj} \sim \frac{1}{\pi} \; \rho \Phi_1 \om t \;  ,
\ee
where $\rho$ is average atomic density. Denoting the effective amplitude
$$
 A \equiv \frac{1}{2} \; \rho \Phi_1 = 2\pi \rho \; \frac{a_1}{m}  \;,
$$
we get the same amplitude-time relation (8) as in the case of the trap
modulation. Thus, these two methods of perturbing condensate should lead
to the same consequences, producing similar strongly excited nonequilibrium
states.

The nature of the possible nonequilibrium states essentially depends on the
atomic scattering length as well as on trap geometry. Important quantities
quantifying these characteristics are the aspect ratio
\be
\label{14}
\al \equiv \frac{\om_z}{\om_\perp} =
\left ( \frac{l_\perp}{l_z} \right )^2 \; ,
\ee
in which $l_\perp \equiv 1/\sqrt{m \omega_\perp}$,
$l_z \equiv 1/\sqrt{m \omega_z}$ are the oscillator lengths, and the effective
coupling parameter
\be
\label{15}
 g \equiv 4\pi N \; \frac{a_s}{l_\perp} \;  .
\ee

It is possible to use other methods for perturbing Bose-Einstein condensate,
for instance, by creating laser-induced obstacles \cite{Allen_24}, which is
equivalent to the perturbation by external potentials, that is, to trap
modulation.

\section{Nonequilibrium atomic states}

The system is initially prepared in an equilibrium state, at temperature
close to zero, with almost all atoms being in Bose-Einstein condensate. Energy
is injected into the system by modulating the trapping potential as described
in Refs. \cite{Henn_14,Shiozaki_15,Seman_16,Bagnato_20}. Injecting energy into
the trap produces a sequence of nonequilibrium states that we have studied
experimentally as well as by computer simulations for the order-function
equation (1), adding there the attenuation by replacing in Eq. (1) $i$ by
$i - \gamma$, which imitates the loss of energy and atoms
\cite{Tsubota_18,Nemirovskii_19}. The experimental results are in very good
agreement with computer simulations. Physical effects, occurring during the
process of perturbing the system can be illustrated by theoretical estimates
that are in close agreement with both experimental results and numerical
simulations.

\vskip 1mm

{\bf Weak perturbation}. Starting pumping into the trap the injected energy
(3), first, creates elementary collective excitations corresponding to small
fluctuations around the equilibrium state. This regime of {\it weak perturbation}
lasts in the interval of energies
\be
\label{16}
 0 < E_{inj} < 2 E_{vor} \;  ,
\ee
where the injected energy is smaller than the energy required for exciting at
least a pair of vortices with opposite vorticities. Such a pair makes the total
vorticity of the atomic cloud zero, as it should be, when the perturbing
potential does not impose rotation. It is worth stressing that the created pair
does not form a bound state, but the vortices are rather independent of each
other. The energy of a vortex per atom can be defined
\cite{Courteille_25,Pethick_26} as
\be
\label{17}
E_{vor} = \frac{0.9\om_\perp}{(\al g)^{2/5} } \; \ln ( 0.8 \al g) \;   .
\ee

Keeping in mind the setup of experiments
\cite{Henn_14,Shiozaki_15,Seman_16,Bagnato_20} with $^{87}$Rb, we have the
trap characteristics
$$
\om_\perp= 1.32 \times 10^3 {\rm s}^{-1} \; , \qquad
\om_z = 1.45 \times 10^2 {\rm s}^{-1} \;   ,
$$
$$
l_\perp= 0.74 \times 10^{-4} {\rm cm}\; , \qquad
l_z = 2.25 \times 10^{-4} {\rm cm}  \qquad ( ^{87}Rb ) \;   ,
$$
which implies the aspect ratio $\alpha = 0.11$. The $^{87}$Rb atoms have the
scattering length $a_s = 0.577 \times 10^{-6}$ cm. With the number of atoms
$N \approx 2 \times 10^5$, the effective coupling (15) is
$g = 1.96 \times 10^4$. This gives the pair vortex energy as
$2 E_{vor} = 0.566 \times 10^{-12}$ eV. It is convenient to measure the
energies in units of the transverse trap energy
$E_\perp \equiv \hbar \omega_\perp$ that in the present case is
$E_\perp = 0.869 \times 10^{-12}$ eV. Then $2 E_{vor} = 0.65 E_\perp$.

In order to stress that the appearance of vortices as well as of other
nonequilibrium states essentially depends on the involved sort of atoms and
on the trap geometry, we compare the typical values for $^{87}$Rb atoms
with the setup employed in the group of Hulet \cite{Pollack_27,Pollack_28}
dealing with $^7$Li atoms and a more elongated trap, where
$$
\om_\perp= 1.48 \times 10^3 {\rm s}^{-1} \; , \qquad
\om_z = 0.304 \times 10^2 {\rm s}^{-1} \;   ,
$$
$$
l_\perp= 2.5 \times 10^{-4} {\rm cm}\; , \qquad
l_z = 1.7 \times 10^{-3} {\rm cm} \qquad ( ^7Li )  \;   ,
$$
which gives the aspect ratio $\alpha = 0.021$, an order smaller than in the
case of $^{87}$Rb. The scattering length of $^7$Li is tuned by Feshbach
resonance to $a_s = 3.2 \times 10^{-8}$ cm, which is two orders shorter than
that of $^{87}Rb$. The effective coupling $g = 0.48 \times 10^3$ is much
smaller than for $^{87}Rb$. The energy, required for vortex generation, is
$2 E_{vor} = 1.45 \times 10^{-12}$ eV. The transverse energy here is
$E_\perp = 0.97 \times 10^{-12}$ eV. Hence $2 E_{vor} = 0.49 E_\perp$. In
this setup, it is more difficult to create vortices, since the required
injected energy $E_{inj}$ should be about three times larger than in the
case of $^{87}$Rb.

In the regime of weak perturbation (16), vortices and other topological modes
are not created, unless the modulating potential is tuned to a transition
frequency $\omega_n \equiv E_n - E_0$, corresponding to a transition between
the ground-state condensate, with energy $E_0$, and a coherent topological
mode \cite{Yukalov_8,Bagnato_20,Yukalov_22,Yukalov_23,Yukalov_29}, with energy
$E_n$. The condition $\omega = \omega_n$, when topological modes are generated,
even under weak perturbation, can be called {\it topological resonance}. Note
that this is not the standard parametric resonance \cite{Bogolubov_30}, when
$\omega = 2 \omega_0$, where $\omega_0$ is a system natural frequency, or
more generally, when $k \omega = 2 \omega_0$, where $k = 1,2, \ldots$. The
natural frequency $\omega_0$ characterizes the motion of the system as a whole
and, for trapped atoms, is defined by the effective oscillator trap frequency.
However the transition frequencies $\omega_n$ describe the internal transitions
between the collective energy levels and, for an interacting system, $\omega_n$
can be very different from the effective oscillator trap frequency
\cite{Yukalov_8,Yukalov_31,Yukalov_32}. Topological modes can also be created
under harmonic generation, when $k \omega = \omega_n$, where $k = 1,2, \ldots$,
under parametric conversion, when there are two external modulating fields,
with the frequencies $\omega_1$ and $\omega_2$ such that
$\omega_1 \pm \omega_2 = \omega_n$, and under combinations of these resonance
conditions. But if there is no a kind of a resonant tuning, the topological
modes do not arise.

When considering nonequilibrium states, it is necessary to distinguish the
atomic cloud itself and the subsystem of its elementary excitations. The atomic
cloud can be slightly perturbed, but, nevertheless, such a perturbation can
result in the creation of many elementary excitations. The waves, corresponding
to these excitations, are involved in wave interactions and are described by
quantum kinetic equations \cite{Gust_33,Reichl_34,Gust_35}. In the subsystem
of these excitations, there can arise weak wave turbulence
\cite{Dyachenko_36,Nazarenko_37}, which is inhomogeneous in the presence of
a trap \cite{Lvov_38}.

\vskip 1mm

{\bf Vortex formation}. As soon as the injected energy surpasses the vortex
energy (17), vortices start being created. The regime of {\it vortex formation}
corresponds to the interval
\be
\label{18}
 2 E_{vor} < E_{inj} < E_{tur} \;  ,
\ee
until the number of vortices is so high that the system turns into turbulent
regime, which happens when the number of vortices reaches a critical value
$N_c$, corresponding to the energy
\be
\label{19}
 E_{tur} = N_c E_{vor} \;  .
\ee
The critical number can be defined by the condition that the mean distance
between the outer parts of the vortices becomes equal to their diameter
$2 \xi$, given by twice the coherence length $\xi \sim \hbar/mc$, where
$c \sim (\hbar/m) \sqrt{4 \pi \rho a_s}$ is the sound velocity. This means
that the distance between the centers of the vortices is $4 \xi$, which gives
\be
\label{20}
 N_c = \left ( \frac{r_\perp}{4\xi} \right )^2 \;  .
\ee
For the experiments with $^{87}$Rb, the atomic cloud radius and length are
$r_\perp \approx 4 \times 10^{-4}$ cm and $L \approx 6 \times 10^{-3}$ cm,
respectively. The coherence length is $\xi \approx 2 \times 10^{-5}$ cm. Thus,
the critical vortex number is $N_c \approx 25$, which is in perfect agreement
with experiments and simulations. This corresponds to the vortex turbulence
energy $E_{tur} = 0.71 \times 10^{-11}$ eV, or in units of the trap energy,
$E_{tur} = 8.2 E_\perp$.

It is much more difficult, if possible at al., to create vortex turbulence
in an elongated trap with trapped $^7$Li, as in the experiments of the Hulet
group \cite{Pollack_27,Pollack_28}. In the latter case, the cloud radius and
length are $r_\perp \approx 3 \times 10^{-4}$ cm and
$L \approx 2 \times 10^{-2}$ cm, while the coherence length is
$\xi \approx 2 \times 10^{-4}$ cm, so that a single vortex would fill almost
the whole trap. Formally, the critical number (20) would be $0.14$. However,
there is no turbulence with a single vortex.

Typical vortex states are illustrated in Fig. 1 showing the results of numerical
calculations for the atomic density in a transverse cross-section of the trap,
for the setup characterized by the experimental data with $^{87}$Rb atoms. The
higher density corresponds to brighter colour. So that vortices are shown as
black spots. By calculating the vorticity, we find that practically all
vortices have vorticity plus or minus one. Typical vortex states observed in
the time-of-flight experiment are presented in Fig. 2. Again, brighter colour
corresponds to higher density.

\begin{figure}[ht]
\begin{center}
\includegraphics[width=0.85\linewidth,keepaspectratio]{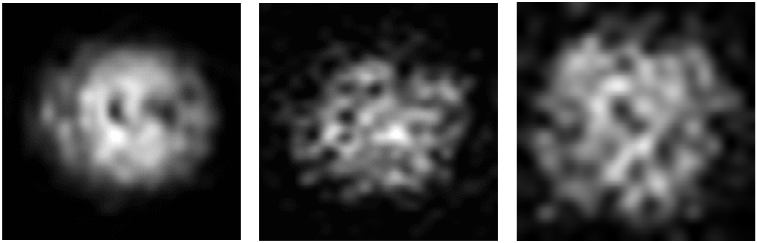}
\end{center}
\caption{(Color online) Vortex state. Typical distributions of the density of
$^{87}$Rb atoms in a transverse cross-section of the trap, found by numerical
simulations. Brighter colour corresponds to higher density. So that vortices
are seen as black spots. }
\label{1}
\end{figure}

\begin{figure}[ht]
\centerline{
\hbox{
\includegraphics[width=3.4cm]{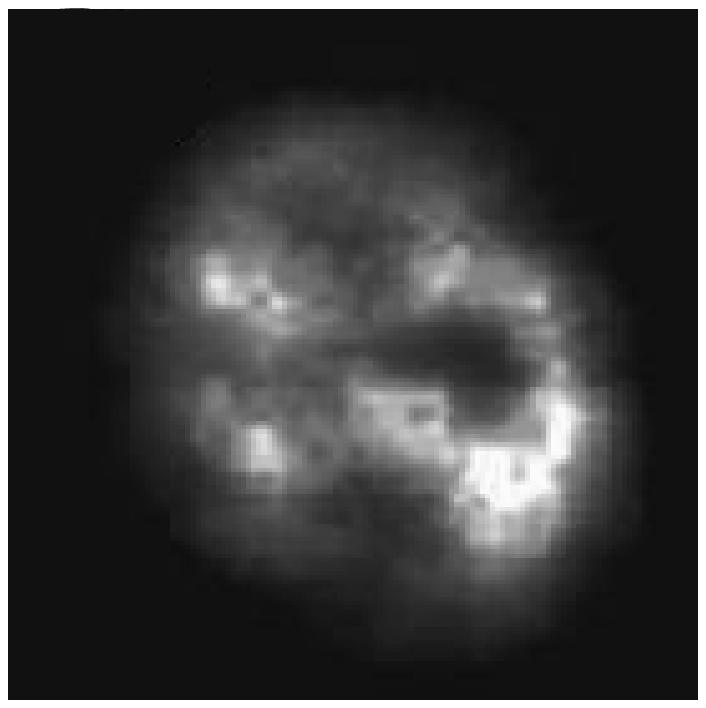} \hspace{2cm}
\includegraphics[width=3.4cm]{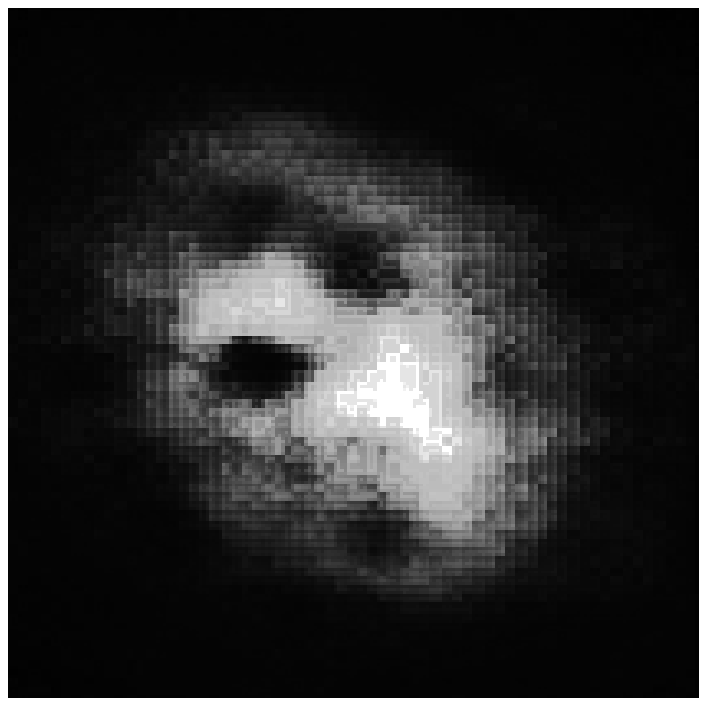} } }
\caption{(Color online) Vortex state. Experimental observation of vortices
in a transverse cross-section of the cloud of $^{87}$Rb atoms in time-of-flight
measurements. Brighter colour corresponds to higher density. So that vortices
are seen as black spots. }
\label{2}
\end{figure}

Increasing the amount of the injected energy produces a larger number of
vortices. Since, as is clear from relation (7), the injected energy is
proportional to the modulation time, the number of vortices, in the vortex
state, increases with time. After their number reaches the critical number 25,
the regime of vortex turbulence comes into play, when the number of vortices
yet increases, but slower than before. This behavior is illustrated by numerical
simulations presented in Fig. 3, showing the number of vortices as a function 
of time, under fixed amplitude. The results are in good agreement with 
experimental data. We see that at some moment of time the number of vortices 
abruptly diminishes. Then the system passes to another regime, termed grain 
turbulence or droplet turbulence, to be described below.

\begin{figure}[ht]
\begin{center}
\includegraphics[width=0.65\linewidth,keepaspectratio]{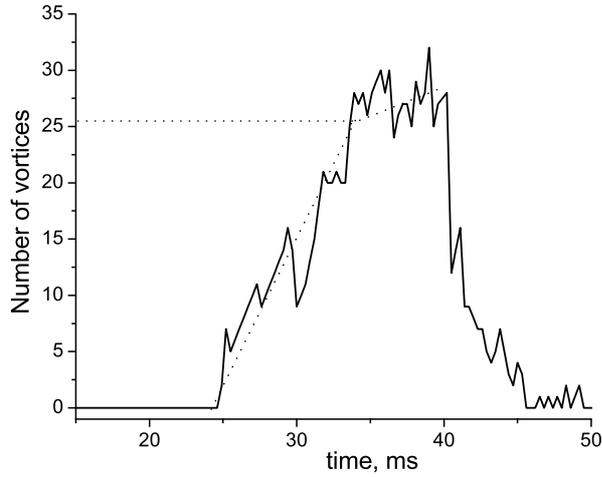}
\end{center}
\caption{(Color online) Number of vortices, found by numerical simulations,
as a function of time, that is, as a function of injected energy. }
\label{3}
\end{figure}

\vskip 1mm

{\bf Vortex turbulence}. The regime of {\it vortex turbulence} lasts till the
number of vortices becomes sufficiently high, so that the mean distance between
vortices is such that their interactions become rather strong, when the
interaction energy of two vortices at the distance $\delta$ from each other
becomes equal to the vortex energy. Equating these energies,
given in Ref. \cite{Pethick_26}, we get
$$
 \ln \left ( 1.4 \; \frac{R}{\xi} \right ) = 2 \ln \frac{R}{\dlt} \;  ,
$$
where $R$ is an effective radius of the cloud. This yields
\be
\label{21}
 \dlt = 1.19 \; \sqrt{R \xi} \;  .
\ee
Taking the Thomas-Fermi radius
$$
  R = l_\perp  \left (  \frac{15}{4\pi}\; \al g \right )^{1/5}
$$
results in the critical distance
\be
\label{22}
 \dlt = 0.86 \; \sqrt{l_\perp \xi} \; (\al g)^{1/10} \; .
\ee
For $^{87}$Rb, the latter is $\delta \sim 0.71 \times 10^{-4}$ cm, so that
$\delta \sim 3.6 \xi$.

One distinguishes {\it random vortex turbulence}, or Vinen turbulence, as
opposed to the {\it correlated vortex turbulence}, or Kolmogorov turbulence
\cite{Nemirovskii_19,Kobayashi_39,Baggaley_40,Baggaley_41,Baggaley_42}. In
Vinen tangles, far-field effects tend to cancel out, the motion of a vortex
line is mainly determined by its local curvature, and the vortex tangle is
homogeneous. The intervortex distance in Vinen turbulence is much larger
than the coherence length. Therefore the summary energy of vortices in the
random Vinen tangle can be approximated by the product of the vortex number
and a single vortex energy.

In the regime of vortex turbulence, we have observed, the distance between
the vortices is yet such that the energy of vortex interactions is smaller
than the energy of a vortex, although interactions can be noticeable. The
vortices are not correlated by the used trap modulation that does not impose
such correlations, nor imposes a rotational polarization. However, to
distinguish precisely what type of turbulence has been realized requires the
study of energy spectra, which is yet in progress. In the present publication,
we concentrate on the study of density distributions that can be directly
observed and which allows for the qualitative classification of the observed
nonequilibrium regimes.

The regime of {\it vortex turbulence} is realized in the interval of the
injected energy
\be
\label{23}
  E_{tur} < E_{inj} < E_{fog} \;   ,
\ee
where the upper energy boundary
\be
\label{24}
E_{fog} = N_c^* E_{vor}
\ee
corresponds to the critical number of vortices, when interactions become
rather strong, which means that the distance between the vortices is smaller
than the critical distance (22). This critical number can be estimated as
\be
\label{25}
 N_c^* = \left ( \frac{r_\perp}{\dlt} \right )^2 \;  .
\ee
For the considered case of $^{87}$Rb, we have $N_c^* = 1.2 N_c \approx 30$.
This defines the energy $E_{fog} = 0.84 \times 10^{-11}$ eV, corresponding
to $E_{fog} = 9.7 E_\perp$.

It is clear that the estimates for the energies $E_{tur}$ and $E_{fog}$ give
their approximate values. It turns out, these values are close to each other.
Although the absolute range between the energies $E_{tur}$ and $E_{fog}$,
of course, is approximate, but its narrowness means that the region of vortex
turbulence is quite narrow. This has really been observed in experiments as
well as in numerical modelling.

The typical cross-section of the random vortex tangle, obtained in numerical
simulations, is demonstrated in Fig. 4. And Fig. 5 shows the vortex turbulent
state observed in the experiment.

\begin{figure}
\begin{center}
\includegraphics[width=0.78\linewidth,keepaspectratio]{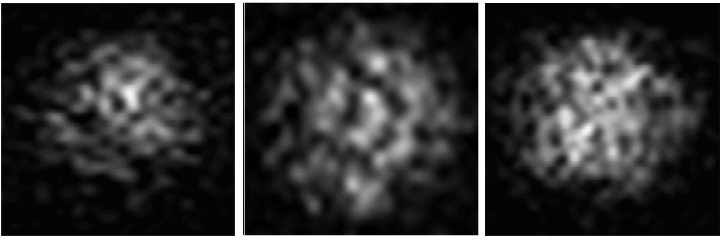}
\end{center}
\caption{(Color online) Vortex turbulence. Cross-section of a random vortex
tangle obtained by numerical simulations. Darker coulor corresponds to lower
density, thus vortices are shown as black spots.  }
\label{4}
\end{figure}

\begin{figure}
\begin{center}
\includegraphics[width=3.4cm]{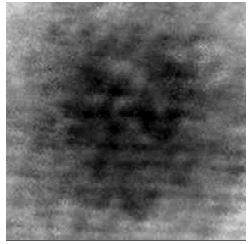}
\end{center}
\caption{(Color online) Vortex turbulence. Experimental observation of a
vortex turbulent state in the cloud $^{87}$Rb atoms. Multiple dark spots
correspond to vortices.  }
\label{5}
\end{figure}

\vskip 1mm

{\bf Grain turbulence}. After the number of vortices in the random tangle
grows to the critical number $N_c^*$, the vortices start strongly interacting
with each other by colliding and destroying each other. Their number sharply
decreases, as is shown in Fig. 3. The remnants of the destroyed vortices form
the pieces of the granulated Bose condensate, which can be called {\it droplets}
or {\it grains}. These droplets, being multiscale in the interval
$(1-5) \times 10^{-5}$ cm, have the typical size $l_{dr} \sim 3 \times 10^{-5}$ cm,
which is close to the coherence length $\xi \sim l_{dr}$, as it should be for
a coherent object. The coherence of a droplet is confirmed by calculating its
phase that is constant through the droplet volume. The droplets are surrounded
by a rarefied gas of much lower density; the ratio of the droplet density to
the density of surrounding $\rho_{dr}/\rho_{sur}$ varies in a wide range: from
a few tens to 100. At each snapshot, the droplets are randomly distributed in
space, forming no spatial structures, reminding water droplets in fog. Therefore,
this regime can be termed {\it droplet turbulence} or {\it grain turbulence}
\cite{Yukalov_43}. The droplets are compact objects, with the linear sizes in
different directions being close to each other. The typical droplet lifetime
is of order $10^{-2}$ s. The droplets do not form any regular structure, since
the average atomic density is not high, being much lower than that necessary
for creating a kind of crystalline order \cite{Vitali_44,Rossi_45}. The random
spatio-temporal distribution of the droplets makes it admissible to call this
state as grain turbulence.

In numerical simulations, considering the opposite process of Bose condensate
formation from an uncondensed gas, the regime of grain turbulence corresponds
to that of strong turbulence \cite{Berloff_46,Zakharov_47}. We prefer to call
it grain turbulence, since this term reflects the physical nature of the system
consisting of dense droplets, or grains, in a rarefied surrounding.

The equivalent interpretation of the granular state could be by imagining it as
a collection of dark solitons, typical of defocusing NLS, inside a Bose-condensed
surrounding. The granular state is similar to the critical balanced state that
is a state where the linear and nonlinear terms are balanced for a wide range of
scales \cite{Proment_48,Proment_49}.

It is worth emphasizing that the transition from one dynamical regime to another
is not an abrupt phase transition, but a gradual crossover, although it can be
very sharp. Therefore the transition lines are, of course, conditional, showing
where one dynamic behavior changes to the other. At the same time, some features
of one dynamic regime can survive to the other region. For instance, in the
regime of grain turbulence, some occasional vortices can occur, although their
number is very small and does not change the overall picture. However, eventually
a few vortices can arise and then annihilate in various locations at random.

Continuing pumping energy into the trapped system destroys Bose condensed
droplets. So that the grain turbulence exists in the interval of energies
\be
\label{26}
 E_{fog} < E_{inj} < E_c \;   ,
\ee
until the injected energy per atom is so high that all Bose condensate becomes 
destroyed, when the energy reaches the value
\be
\label{27}
E_c = k_B T_c \;   ,
\ee
where $T_c$ is the critical temperature of Bose-Einstein condensation. For
the treated case of $^{87}$Rb, the critical temperature is
$T_c = 2.76 \times 10^{-7}$ K. Then we have $E_c = 0.238 \times 10^{-10}$ eV.
This gives $E_c = 27.4 E_\perp$.

Figure 6 presents the cross-section of the atomic cloud in the regime of grain
turbulence, found numerically. Droplets are shown as bright spots surrounded
by dark rarefied regions. Each droplet is formed by Bose-condensed atoms, which
is confirmed by a constant phase inside a droplet. In Fig. 7, an experimentally
observed granular state is shown.

\begin{figure}
\begin{center}
\includegraphics[width=0.78\linewidth,keepaspectratio]{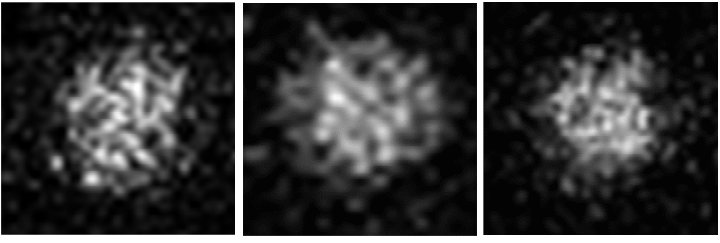}
\end{center}
\caption{(Color online) Grain turbulence. Cross-section of the atomic cloud
composed of dense Bose-condensed grains, randomly distributed in space, inside
a rarefied surrounding. Droplets are seen as bright spots.  }
\label{6}
\end{figure}

\begin{figure}
\begin{center}
\includegraphics[
width=3.4cm]{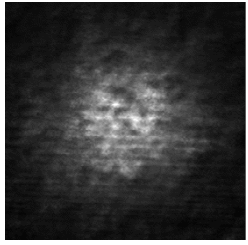}
\end{center}
\caption{(Color online) Grain turbulence. Experimental observation of a
granular turbulent state in the cloud of $^{87}$Rb atoms. Brighter spots
correspond to higher density, while darker regions, to rarefied density.
Droplets are seen as bright spots inside a rarefied surrounding in dark.  }
\label{7}
\end{figure}

\vskip 1mm

{\bf Wave turbulence}. Increasing the amount of injected energy should finally
lead to a complete destruction of condensate \cite{Adhikari_50}. This happens
after the injected energy surpasses the critical
energy $E_c$,
\be
\label{28}
  E_{inj} > E_c \; ,
\ee
when all Bose-Einstein condensate is getting destroyed. Then the system is 
represented by small-amplitude waves, because of which it is termed 
{\it wave turbulence}, or {\it weak turbulence}. The linear sizes of waves 
are in the interval $(0.5 - 1.5) \times 10^{-4}$ cm, so that the typical wave 
size is $l_w \sim 10^{-4}$ cm. The wave density is only slightly greater than 
that of their surrounding, $\rho_w / \rho_{sur} \sim 3$. The phase inside a 
wave, as well as between the waves, is completely random. The system kinetic 
energy is much higher than the interaction energy, so that the system can be
represented as a collection of almost independent modes described by a kinetic
equation with four-wave processes \cite{Nazarenko_37}. The dynamic transition
from grain turbulence to wave turbulence is a crossover, such that in a wide
region grains coexist with waves.

Also, in the weak turbulence regime, there exist  the so-called 
{\it ghost vortices} that are nodal points of the wave field, which are present 
in abundance if the waves are linear \cite{Nazarenko_37}.

The cross-section of the trapped cloud in the regime of wave turbulence is
shown in Fig. 8, where the system density and phase are presented. This figure
is obtained by numerical simulations. Because of the large amount of injected 
energy, required for creating this regime, it has not yet been reached in 
experiments.

\begin{figure}
\begin{center}
\includegraphics[width=0.6\linewidth,keepaspectratio]{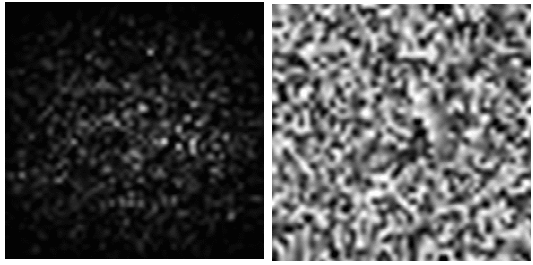}
\end{center}
\caption{(Color online) Wave turbulence. Cross-section of the trapped cloud
in the regime of wave turbulence, found numerically. The left figure shows
the density distribution, where brighter colour corresponds to higher density.
The right figure shows the spatial phase distribution, where the larger phase
is of brighter colour.  }
\label{8}
\end{figure}

To stress the principal difference between the regimes of grain turbulence
and wave turbulence, their characteristic features are compared in Fig. 9.

\begin{figure}
\begin{center}
\includegraphics[width=0.55\linewidth,keepaspectratio]{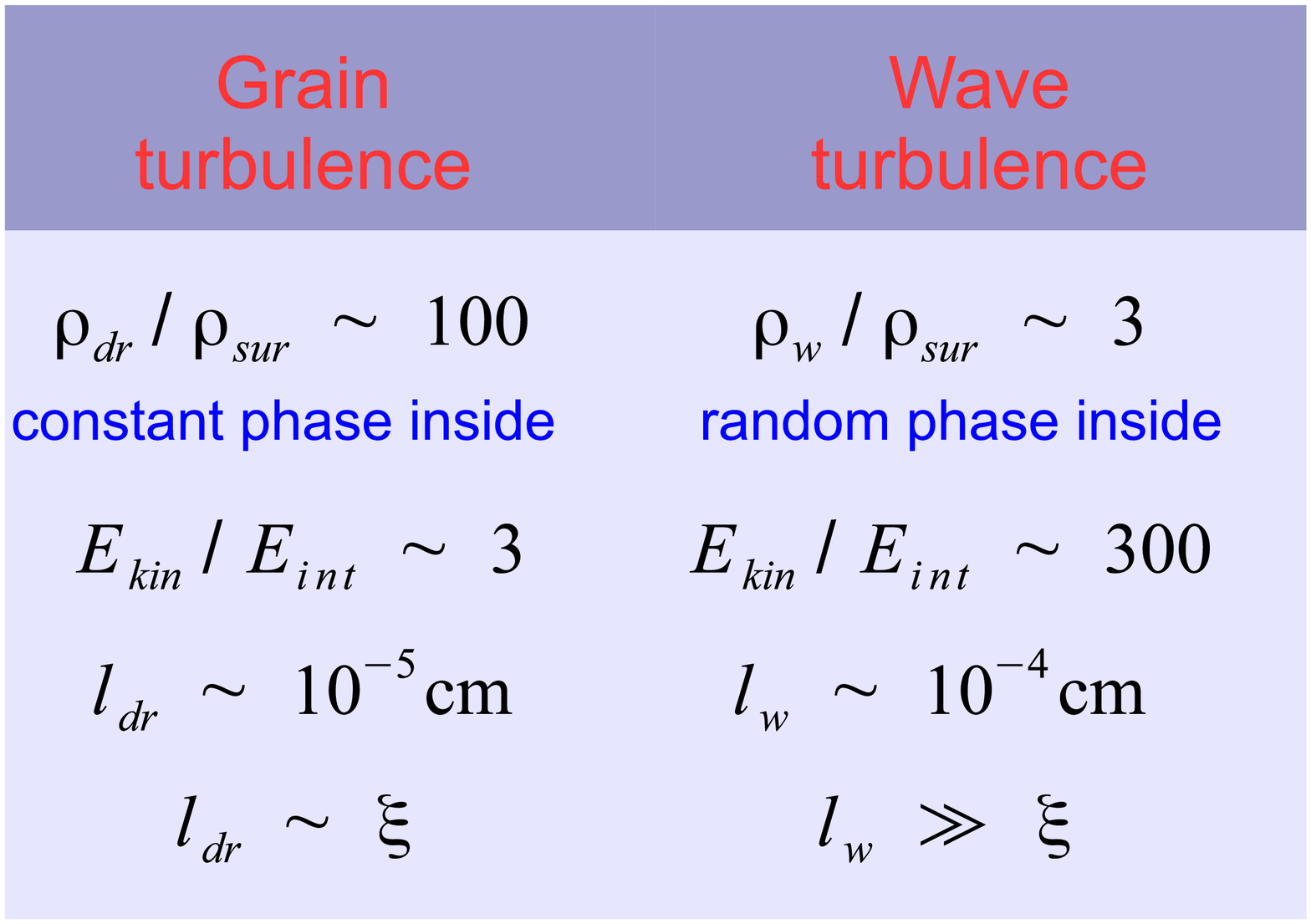}
\end{center}
\caption{(Color online) Characteristic features of the regimes of grain
turbulence and wave turbulence.
  }
\label{9}
\end{figure}

\begin{figure}
\begin{center}
\includegraphics[width=0.57\linewidth,keepaspectratio]{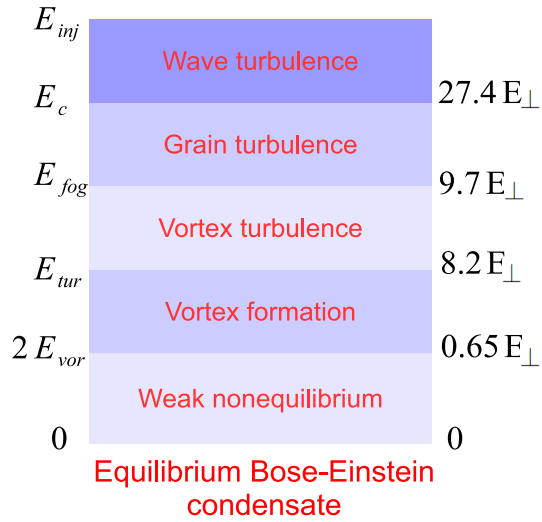}
\end{center}
\caption{(Color online)  Sequence of nonequilibrium states produced by
modulating the trapping potential for $^{87}$Rb atoms.  }
\label{10}
\end{figure}

\begin{figure}
\begin{center}
\includegraphics[width=0.65\linewidth,keepaspectratio]{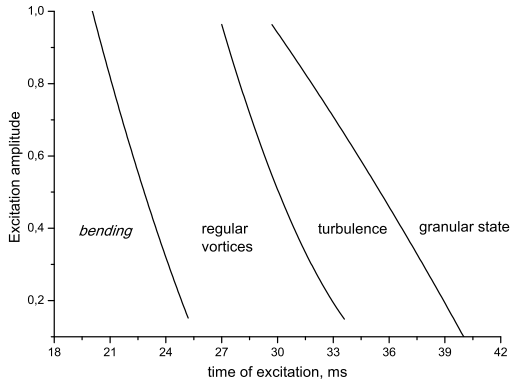}
\end{center}
\caption{(Color online) Numerical amplitude-time phase diagram for $^{87}$Rb.  }
\label{11}
\end{figure}

In conclusion, we have studied both experimentally and by means of computer
simulations strongly nonequilibrium states of trapped $^{87}$Rb atoms, prepared
in Bose-condensed state and subject to strong perturbations realized by
modulating the trapping potential. The sequence of the nonequilibrium regimes,
obtained by computer simulations, is summarized in Fig. 10. And the numerical
amplitude-time diagram is presented in Fig. 11. The latter is in good agreement
with the experimentally found amplitude-time diagram discussed earlier
\cite{Shiozaki_15,Seman_16}. It is interesting to note that the sequence of the
nonequilibrium states, generated in the process of strong perturbation, which
includes weakly nonequilibrium state, vortex state, vortex turbulence, grain
turbulence, and wave turbulence, repeats the analogous states, although in the 
reverse order, as those occurring during the process of equilibration 
\cite{Yukalov_51,Polkovnikov_52} through a nonequilibrium Bose-condensation 
phase transition from a strongly nonequilibrium uncondensed state to an 
equilibrium condensed state \cite{Nazarenko_37,Berloff_46,Zakharov_47,Semikoz_53}. 
Thus, we generate the states typical of the Kibble-Zurek mechanism 
\cite{Kibble_54,Zurek_55}, but in the opposite temporal order. More discussions 
on such an inverse Kibble-Zurek scenario will be given in a separate paper.

\begin{acknowledgements}

We acknowledge financial support of FAPESP (Brazil)
and RFBR 14-02-00723 (Russia). One of the authors (V.I.Y.) is grateful to
M. Tsubota and E.P. Yukalova for discussions.

\end{acknowledgements}

\end{document}